**From boiling point to glass transition temperature: transport coefficients in molecular liquids follow three-parameter scaling**

B. Schmidtke, N. Petzold, R. Kahlau, M. Hofmann, E.A. Rössler*

Universität Bayreuth, Experimentalphysik II, D-95440 Bayreuth, Germany

*) corresponding author

## Abstract

The phenomenon of the glass transition is an unresolved problem of condensed matter physics. Its prominent feature, the super-Arrhenius temperature dependence of the transport coefficients remains a challenge to be described over the full temperature range. For a series of molecular glass formers, we combined $\tau(T)$ from dielectric spectroscopy and dynamic light scattering covering the range $10^{-12}$ s< $\tau(T)$ <$10^2$s. Describing the dynamics in terms of an activation energy $E(T)$, we distinguish a high-temperature regime characterized by an Arrhenius law with a constant activation energy $E_\infty$ and a low-temperature regime for which $E_{coop}(T) \equiv E(T) - E_\infty$ increases while cooling. A two-parameter scaling is introduced, specifically $E_{coop}(T)/E_\infty = f[\lambda(T/T_A\text{-}1)]$, where $f$ is an exponential function, $\lambda$ a dimensionless parameter, and $T_A$ a reference temperature proportional to $E_\infty$. In order to describe $\tau(T)$, in addition, the attempt time $\tau_\infty$ has to be specified. Thus, a single interaction parameter $E_\infty$ extracted from the high-temperature regime together with $\lambda$ controls the temperature dependence of low-temperature cooperative dynamics.

## I Introduction

Although of fundamental importance and extensively investigated the glass transition phenomenon is far from being understood. Its most prominent feature is the super-Arrhenius temperature dependence of transport coefficients such as viscosity or correlation time $\tau$ which is observed when a liquid is strongly cooled and crystallization is avoided. While a simple (molecular) liquid well above its melting point exhibits a viscosity on the order of $10^{-3}$ Pas, upon super-cooling it may finally reach values of $10^{12}$ Pas which are typical of solid body. The corresponding temperature is called the glass transition temperature $T_g$. The slowing-down of dynamics is accompanied by only a weak change in structure. This has lead to the interpretation that the glass transition is a kinetic transition and several theoretical approaches have been developed, yet none is broadly accepted [1,2,3,4]. For example, it remains a great challenge of any theory of the liquid state to provide an interpolation of $\tau(T)$ which covers the full range from the boiling point down to $T_g$.

Often the empirical Vogel-Fulcher-Tammann formula (VFT), $\lg \tau / \tau_0 = D/(T - T_0)$, is applied to fit experimental data. One of the problems one faces when applying VFT is that its parameters depend strongly on the fitting interval and it fails when relaxation data well above the melting point are included. Regarding the divergence of the correlation time implied by VFT at $T_0 < T_g$ doubts have also been raised [5]. Numerous further formulae have been proposed attempting to fit $\tau(T)$ but none is fully satisfying. Another route of searching for a universal temperature dependence or "corresponding states" of liquids relies on scaling, say, the low-temperature regime by introducing some crossover temperature [6,7,8,9]. Yet, in the different approaches the physical meaning of the crossover temperature is quite different, and it is difficult to extract unambiguously the crossover temperature.

Inspecting the experimental situation it turns out that although extensively studied close to $T_g$ molecular glass formers are not sufficiently well investigated in the high temperature regime. With a few exceptions most tests of interpolation formulae for $\tau(T)$ are restricted to time constants above, say, $10^{-9}$s actually ignoring a temperature range of up to 300 K until correlation times on the order of the high-temperature limit $\tau_\infty \cong 10^{-12}$s are reached. The reason for this is that most dielectric relaxation experiments [10,11,12], actually the most popular approach probing molecular reorientation associated with the slowing-down of structural relaxation, do not cover frequencies above a few GHz. Correlation times down to $10^{-12}$s are now easily available when glass-formers are studied by dynamic light scattering (LS) using tandem-Fabry-Perot interferometer and double monochromator [13,14,15,16,17]. We have combined LS data measured up to 440 K of a series of 17 molecular liquids with the data obtained by dielectric spectroscopy and thus cover the entire temperature range needed to attempt a complete description of $\tau(T)$, *i.e.*, which includes both the high- as well as the low-temperature regime of molecular liquids which easily can be super-cooled. Looking for a minimal number of system-specific parameters controlling $\tau(T)$ in the range $10^{-12}$ - $10^{2}$s we will show that actually three parameters are sufficient to achieve this task.

Figure 1(a) presents in an Arrhenius representation dielectric correlation times (open symbols) [10,18,19,20,21,22] together with few literature data [23,24,25]; in addition we have included our new together with previously published LS data [14,15,16,17] (full symbols). In the cases of dimethyl phthalate (DMP) and m-tricresyl phosphate (m-TCP) we only use LS data including our photon correlation spectroscopy data. It is obvious that adding the LS data extends significantly the temperature range to be included in a full scale description of $\tau(T)$, a fact better seen when the data are plotted as function of temperature (Fig. 1(b)). Even including the LS data, however, only in the case of the low-$T_g$ liquids (say, $T_g$ < 180 K) one reaches correlation times on the order of $10^{-12}$s at 440 K. For the systems with high $T_g$ this limit is not reached. In the case of o-terphenyl ($T_g$= 245 K) viscosity data [26] are available being measured up to almost 700 K which allows to cover the high-temperature regime also for this high-$T_g$ system (crosses). Here, with regard to our LS data measured up to 440 K about another 260 K have to be covered to finally reach $10^{-12}$s. In our analysis we also included viscosity data of ααβ-trisnaphthyl benzene (TNB; $T_g$=343 K) [27]. As different rank reorientational correlation functions are probed by DS and LS, respectively, one expects some small difference in the absolute values of $\tau(T)$ which, however, can be neglected on a logarithmic scale.

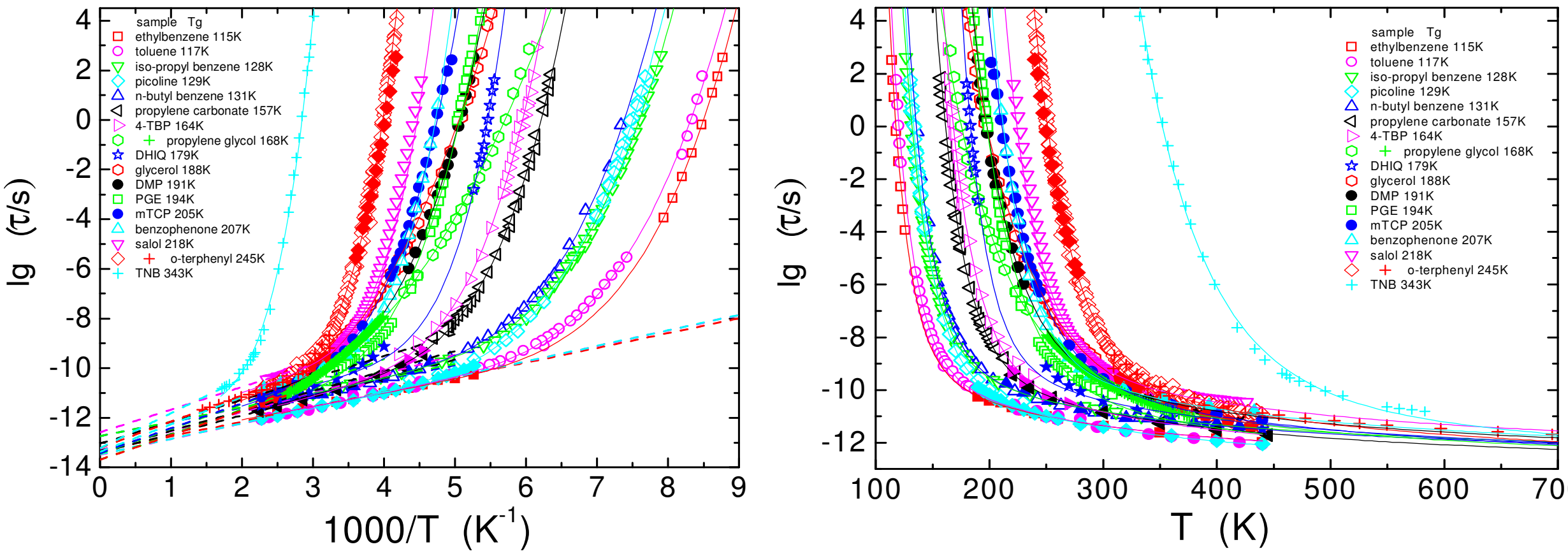

Figure 1: Reorientational correlation times of molecular liquids ($T_g$ indicated) obtained by dielectric spectroscopy (open symbols) [10,18,19,20,21] and dynamic light scattering (full symbols) (this work and [14,15,16,17]); 4-TBP: 4-tert-butyl pyridine, DHIQ: decahydro-isoquinoline; DMP: dimethyl phthalate, PGE: monoepoxide phenyl glycidyl ether, m-TCP: m-tricresyl phosphate; data for n-butyl benzene from [28,29], iso-propylene benzene from [30,31,32]; viscosity data for o-terphenyl [26], trisnaphthyl benzene (TNB) [27] and propylene glycol [33] (crosses); straight dashed lines: high-temperature Arrhenius behavior; solid lines: full fit by eq. (1) and (4).

It is well known from transport data in low-viscosity (non-glass-forming) liquids that their temperature dependence is described by an Arrhenius law (34). This is also seen from inspecting the data in Fig. 1. Only at low temperatures a simple Arrhenius law, $\lg \tau / \tau_\infty = E_\infty / T$, fails and the apparent activation energy $E(T)$ strongly increases. For example, in the case of o-terphenyl the high-temperature regime with $E(T) = E_\infty$ = const. ends around 500 K. Although the energy $E_\infty$ is an apparent quantity and must not be connected to some single-particle barrier in the liquid we take the Arrhenius high-temperature dependence as an empirical fact and as a starting point of our analysis. Explicitly, we assume that $\tau(T)$ can be described by the following expression

$$\tau(T) = \tau_\infty \exp((E_\infty + E_{coop}(T))/T) \tag{1}$$

where the apparent activation energy $E(T)$ (in Kelvin) is decomposed into a temperature independent part $E_\infty$ and a temperature dependent part $E_{\mathrm{coop}}(T)$. The quantity $E_{\mathrm{coop}}(T)$ reflects the cooperative dynamics becoming dominant at low temperature. As will be demonstrated we claim that a scaling applies for $E_{\mathrm{coop}}(T)$; explicitly

$$E_{coop}(T)/E_\infty = f[\lambda(T/T_A - 1)] \tag{2}$$

where $f$ is a universal function, $T_A$ a reference temperature which presumably is proportional to $E_\infty$, and $\lambda$ a dimensionless system-specific parameter. In other words, the quantity $E_{\mathrm{coop}}(T)$ which reflects the low-temperature cooperative dynamics is itself a function of $E_\infty$. We note that such scaling has already been proposed by theoretical [3,4,35] as well as experimental approaches [36]. Yet, up to our knowledge none made a systematic study on several liquids including high-temperature data which as mentioned have been rare.

In Fig. 2, by plotting $T \lg(\tau/\tau_\infty) - E_\infty$ the quantity $E_{\mathrm{coop}}(T)$ is displayed as a function of temperature. The high-temperature regime is now characterized by $E_{\mathrm{coop}}$ being essentially zero while at low temperatures $E_{\mathrm{coop}}(T)$ strongly increases. The quantity $E_{\mathrm{coop}}(T = T_g)$, *i.e.*, the energy at the reference temperature $T_g$ is the higher the higher is $T_g$, a trend already anticipated in Fig. 1. Indeed, the ratio $E_{\mathrm{coop}}(T_g)/T_g$ appears to be constant (*cf.* Fig. 3). Assuming strict proportionality between $E_{\mathrm{coop}}(T)$ and $T_g$ would suggest a two-parameter scaling of the kind

$$(T\,\lg(\tau\ /\tau_\infty)-E_\infty)/T_g = E_{coop}(T)/T_g \equiv f[\lambda(T/T_g-1)] \qquad (3)$$

Yet, the scatter observed in the ratio $E_{coop}(T_g)/T_g$ is actually significant. We take this as a hint that $T_g$ is not the correct reference temperature needed to scale the data. Of course, this is expected since $T_g$ is an "isodynamic point" chosen arbitrarily. One may further speculate whether $E_\infty \propto T_g$ holds. This is also checked in Fig. 3. Indeed both the ratios $E_{coop}(T_g)/T_g$ and $E_\infty/T_g$ appear to be constant although scatter is observed. We end up with the important prospect that the temperature dependence of the low-temperature dynamics may be linked to the high-temperature activation energy $E_\infty$.

Proportionality between $E_\infty$ and $T_g$ would suggest to choose as new reference temperature $T_A \equiv E_\infty$. However, this choice is not possible as then in particular the non-fragile liquids glycerol or propylene glycol cannot be mapped to a master curve. Again, we conclude that a strict proportionality between $E_\infty$ and $T_g$ does actually not hold. Thus, we searched for a choice of the reference temperature $T_A$ for which scaling works best. This was done by varying $T_A$ in the range $T_g < T_A < E_\infty$. Furthermore, we exploit the fact that $E_{coop}(T)$ is essentially an exponential function of temperature as is demonstrated in Fig. 2(b). Straight lines are observed for the low-$T_g$ systems. In the case of the high-$T_g$ systems and particularly for the non-fragile liquids the curves bent over at low values of $E_{coop}$. Most probably this is due to an underestimated $E_\infty$. This once again points to the principal difficulty of determining $E_\infty$ correctly in the case of high-$T_g$ liquids. Moreover, we are faced with the problem to analyze $E_{coop}(T)$ containing the error of a not correctly chosen $E_\infty$ (along Eq. 1) in addition to scatter reflecting experimental errors in $\tau(T)$. Hence, we take recourse to some optimization procedure.

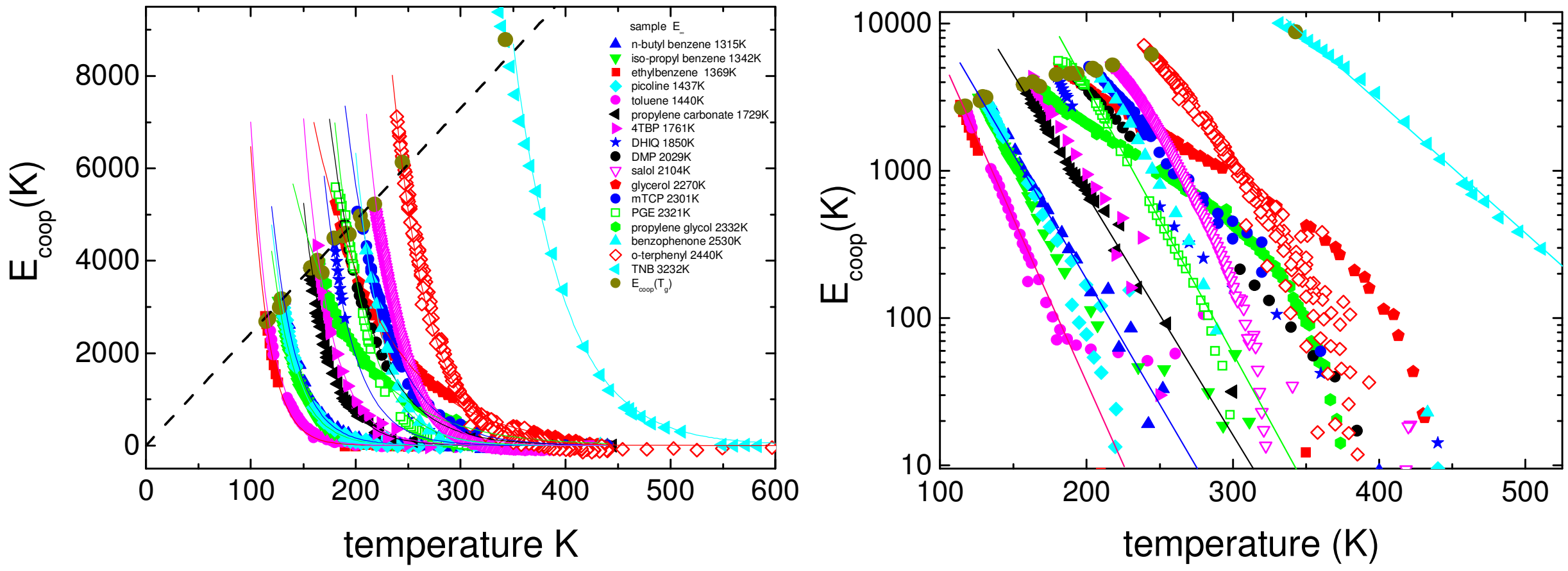

Fig. 2: (a) The quantity $E_{coop}(T)$ (*cf.* eq. 1) as a function of temperature (symbols like in Fig. 1); dashed line connecting $E_{coop}(T=T_g)$; (b) data in a semi-logarithmic plot; straight lines signal exponential dependence in particular for low-$T_g$ liquids. For high-$T_g$ and non-fragile liquids the curves bent over at low $E_{coop}$ values probably due to an underestimated value of $E_\infty$.

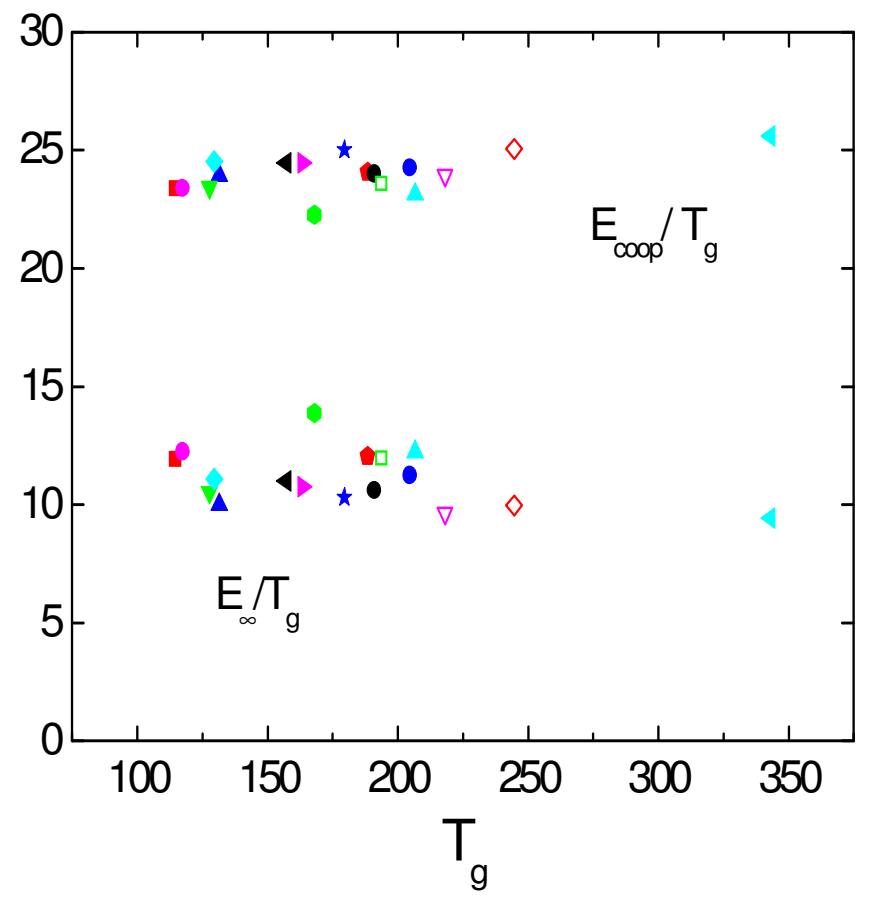

Fig. 3: Correlation between $E_{coop}(T=T_g)$ and $E_\infty$, respectively, with the glass transition temperature $T_g$.

We fit the $E_{coop}(T)$ in Fig. 2(a) for all systems by the expression

$$E_{coop} / E_\infty = a \exp[-\lambda(\frac{T}{bE_\infty} - 1)] \qquad (4)$$

where $a$ and $b$ are universal (global) parameters to be determined under the condition that the correlation between the experimental and fitted value of $E_\infty$ (by applying eq. 3 to the data in Fig. 2(a)) becomes best. Our search yields the result that $E_{coop}(T_A) = E_\infty$ ($a \cong 1$) and $T_A = 0.104 E_\infty$ ($b \cong 0.104$)., the inset in Fig. 4(b) shows a satisfying correlation between the fitted and the experimental value of $E_\infty$. In Fig. 4(a) we show $E_{coop}(T)/E_\infty$ vs. $T/E_\infty$, and in Fig. 4(b) the master curve is displayed semi-logarithmically. The obtained values of $T_A$ and $\lambda$ are included in Fig. 4(a) and 4(b), respectively. In Fig. 1(a) and 1(b) very satisfying three-parameter ($E_\infty$,$\lambda$,$\tau_\infty$) fits of $\tau(T)$ by Eq. (1) and (4) are shown which cover all the available data from essentially the boiling point down to $T_g$; a much larger temperature interval can be covered compared to VFT. These good fits may also be taken as a crosscheck to demonstrate that indeed in most cases no large errors are involved in determining $E_\infty$.

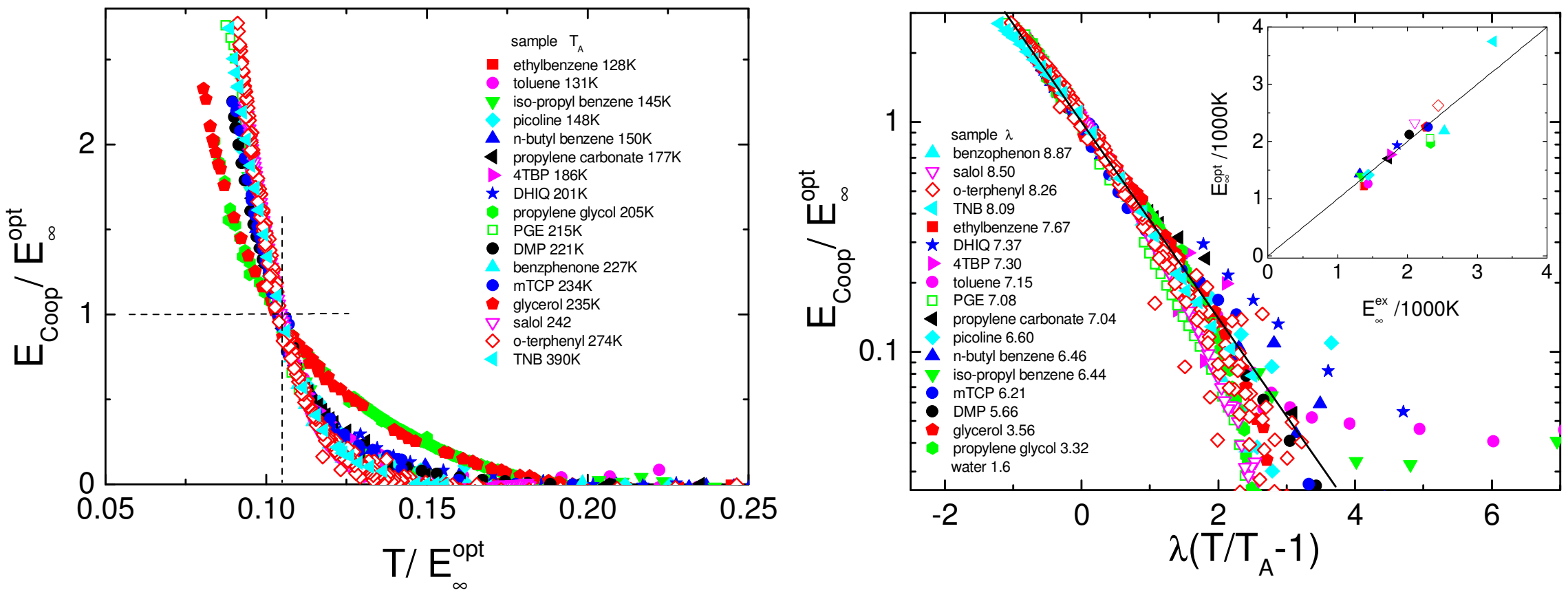

Fig. 4: (a) $E_{coop}/E^{opt}_\infty$ as a function of the reduced temperature $T/E^{opt}_\infty$ with $E^{opt}_\infty$ obtained by the optimization strategy; (b) corresponding master curve obtained by introducing the dimensionless fragility parameter $\lambda$ and $T_A$= 0.104 $E^{opt}_\infty$.

Concluding we propose a three-parameter interpolation of the complete temperature dependence of transport quantities in molecular liquids which are easily super-cooled, i.e., when time constants in the range of $10^{-12}$- $10^{2}$s are covered. We note that one has to exclude diffusion data from this scaling as they show a “decoupling phenomenon” close to $T_g$ [37]. The decomposition along eq. (1) is not unique and our sole justification is the success of the corresponding scaling, a minimal set of system-specific parameters, and furthermore, a simple exponential describes of $E_{coop}(T)$. The quantity $E_\infty$ extracted from the high-temperature transport data we interpret as an interaction parameter which together with the dimensionless parameter $\lambda$ controls the low-temperature behavior of $\tau(T)$.

Recently, Capaccioli and Ngai [38] reiterated the controversy of providing a reliable estimate of $T_g$ of water. They suggested $T_g$ =136 K as the best value. We fitted our formula to the data from [39], which Capaccioli and Ngai also used. So we are able to extract the fragility parameter $\lambda$ = 2.6 (referring to a steepness index $m$ = 37) which is close to that of glycerol and propylene glycol both of which are non-fragile. As expected water is a hydrogen network forming liquid similar to glycerol and thus is not a fragile glass former. Actually, Capaccioli and Ngai estimated $m$ = 44 which is in good agreement with our prediction.

All in all, the present finding is of great relevance for future theory of the glass transition phenomenon associated with the super-Arrhenius temperature dependence of the correlation time which sets in well above the melting point and thus is an important feature of any condensed matter.

Acknowledgement

The authors thank D. Kruk, A. Bourdick, and B. Pötzschner for helpful discussions, and financial support of Deutsche Forschungsgemeinschaft (DFG) through project RO 907/11 and RO 907/15 is appreciated.